\documentclass[prl,twocolumn,floatfix]{revtex4-1}
\usepackage{amsmath,amssymb,latexsym} 
\usepackage{rotating}

\usepackage{graphicx}
\usepackage{color}

\begin{document}

\title
{Liquid-liquid Phase transitions in  silicon.
}

\author
{
 M.W.C. Dharma-wardana}
\email[Email address:\ ]{chandre.dharma-wardana@nrc-cnrc.gc.ca}
\affiliation{
National Research Council of Canada, Ottawa, Canada, K1A 0R6
}

\author{Dennis D. Klug}\email[Email address:\ ]{dennis.klug@nrc-cnrc.gc.ca}
\affiliation{
National Research Council of Canada, Ottawa, Canada, K1A 0R6
}

\author
{Richard C.  Remsing,}
\affiliation {Rutgers University,
Department of Chemistry and Chemical Biology, Piscataway, NJ 08854-8019 USA
}

\begin{abstract}
We use computationally simple neutral  pseudo-atom (`average atom' or 'single-center') 
density functional theory (DFT) as well as  standard N-atom  DFT  to elucidate liquid-liquid
 phase transitions (LPTs) in  supercooled liquid silicon at 1200K, as well as in  
 silicon as 'warm dense matter' at 11604K (1 eV). An 
ionization-driven transition  and three LPTs
including the known LPT near
2.5 g/cm$^3$ are found. They are robust even to 1 eV. 
The  pair distributions functions, pair potentials,
 electrical  conductivities, and compressibilites
 are reported. The origin of the LPTs are clarified  within 
 a Fermi liquid picture  of strong electron-ion
scattering at the Fermi energy and complements the 
commonly used transient covalent bonding picture.
\end{abstract}

%
\maketitle
{\it Introduction}--
Light elements like C, Si, P etc., are  insulators or semiconductors
that become dense metals  when molten. They manifest
transient covalent binding ({\it tr-cb}) even
after melting~\cite{Si-Aptekar79,Stich89, DWP-carb90, OkadaSi12,CPP-carb18}. 
Warm-dense matter (WDM) 
techniques~\cite{McBride-Si-19,Ng05} can be used to
study these materials over a broad density ($\bar{\rho}$) and
 temperature ($T$) range~\cite{sxhu2017}.
A recent study of WDM carbon
provided pair-distribution functions (PDFs) $g(r)$ and other data
 suggestive of a  phase transition from a highly correlated WDM state
 to a weakly  correlated plasma, driven by a change
 in ionization~\cite{CPP-carb18}.
The  {\em supercooled} liquid silicon ($l$-Si) near 1200K, 
 polymorphic forms of silicon, as well as model fluids
 have been studied for a liquid-liquid phase transition (LPT)
between a high-density liquid  (HDL)  and a low-density
liquid  (LDL)~\cite{SastryAngel03,Ashwin04,McMillan05,Beaucage05,
Daisen07SiPhDia,GaneshSiLPPT-09,Baye10,Vasisht11}. 
 Remsing {\it et al. }\cite{Remsing17,Remsing18}
confirmed the LPT
via  density functional theory (DFT)
based molecular dynamics
(MD) for $l$-Si, using the
 `strongly constrained and appropriately normed' (SCAN)
 XC-functional~\cite{SCAN2013} suitable for
{\it tr-cb} systems. The non-metallic LDL ($nm$-LDL) is
less dense than the solid.
Here we examine Si over a  range of $\bar{\rho},T$  and find 
three LPTs viz., LPT2 near 2 g/cm$^3$, LPT2.5 near 2.27-2.57 g/cm$^3$, 
LPT3 near 3 g/cm$^3$ and  an
 ionization-driven transition (IDT) at 1.5 g/cm$^3$. 
The IDT and the LPTs are found to be robust and
may be studied even at higher $T$.

{\it Method}-- we use standard $N$-center DFT-MD  and one-center
 neutral pseudoatom (NPA) methods.
 The NPA reduces {\it both} the electron-electron and ion-ion
  manybody problems to two coupled one body problems
 via exchange-correlation
 functionals~\cite{DWP82,ilciacco93,eos95,CDW-N-rep19}.
 The NPA and  the hyper-netted-chain (HNC) equation provide rapid,
 DFT results via mere `laptop' calculations
 for  $g(r)$, the structure factor $S(k)$, thermodynamic and transport properties
(see supplemental matter,
 SM~\cite{supmat}).

\begin{figure}[t]                    
\includegraphics[width=0.96\columnwidth]{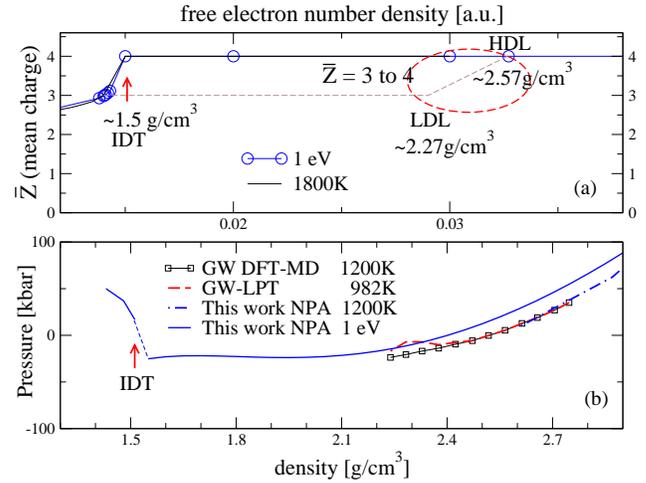}
 \caption
{(online color) (a) The Si charge  ${\bar{Z}}$
versus the free electron density $n_f$. At low density,
a drop in $\bar{Z}$  may cause an Ionization Driven Transition (IDT).
The ellipse indicates the HDL-LDL LPT.
(b) The HDL branch of the Ganesh-Widom (GW)~\cite{GaneshSiLPPT-09}
 pressure and the NPA  at 1200K agree. The pressure at 1 eV (11,604 K)
 is discontinuous
at the IDT.}
\label{DenZbar.fig}
\end{figure}

{\it Phase transitions}-- In DFT-MD the  free energy $F(\bar{\rho},T)$
 is calculated using an $N$-atom  simulation cell, with
 $N\sim 100-500$. The NPA  uses $N=1$ and inputs the free electron density $\bar{n}$
 and $T$ to construct the equilibrium ionic  density $\bar{\rho}$,
the mean ionic charge  $\bar{Z}$,  ion-electron and ion-ion pair-potentials.
 The free energies and linear transport properties 
(e.g, conductivity $\sigma$)  are obtained using only
 NPA-generated quantities~\cite{eos95,supmat}.
 In DFT-MD,
 the Kubo-Greenwood (KG) dynamic conductivity $\sigma(\omega)$ is calculated
 and averaged over many fixed ion configurations.  A  mean-free path 
 and a Drude model are invoked  by KG to get 
 the static KG conductivity $\sigma(\omega\to0)$.

Several mechanisms for phase transitions in $l$-Si exist.
1. In $l$-Si ($\bar{\rho}$=2.56 g/cm$^3$)  near the melting
 point (1683K), $\bar{Z}$
is four. On lowering $\bar{\rho}$  sufficiently, $\bar{Z}$ 
drops,  and  phase transitions may occur (Fig.~\ref{DenZbar.fig}).
The transition of $\bar{Z}$ from 4 to 3 occurs at $\rho<$1.5 g/cm$^3$ for
 $l$-Si at 1200K. Such  wide ranging $\bar{\rho},T$ studies cannot be
done  using model-potentials (e.g., Stillinger-Weber)~\cite{ZhaoLqdSi16}
as the role of the electron quantum fluid is suppressed.

2. If the first peak of  $S(k)$  for some density range falls near 2$k_F$
in a metallic fluid (as in $l$-Si), then concerted
scattering across the Fermi surface causes a giant Kohn anomaly that translates 
into {\it tr-cb}  with lifetimes typical  of
 phonon vibrations~\cite{DWP-carb90}.  That is  the ``Fermi liquid'' picture  of {\it tr-cb} 
 formation. The presence of {\it tr-cb}  splits the main peak of $S(k)$ and causes
 a peak in the PDF close to the  Si-Si bond distance $r_b$.

3. In  simple metallic fluids, e.g., $l$-Al, the first
peak of  $g(r)$ occurs  near $r_1$ $\sim$1.6$r_{ws}$,
 where the Wigner-Seitz  radius $r_{ws}$ is $\{3/(4\pi\bar{\rho})\}^{1/3}$.
 This is  a hard-sphere packing effect acting against the
electron cohesive energy.
Complex fluids  can lower $F$  further
 if $\bar{\rho}$  adjusts via an LPT to bring $r_1$ 
near   $r_b\sim$2.1-2.5 \AA \ 
for  $l$-Si, and $\sim 1.5$ \AA \ for
 $l$-C. Transient bonding increases the
 available configurations and entropy, lowering
$F$ to drive the LPT~\cite{ZhaoLqdSi16}. 

\begin{figure}[t]    
\includegraphics[width=0.98\columnwidth]{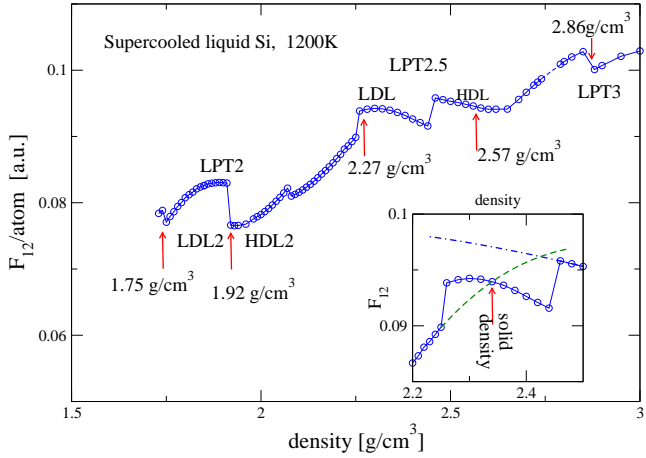}
 \caption
{(online color)The ion-ion part of the  free energy $F_{12}$
shows  discontinuities at possible
LPTs. The LPT2.5 at 2.27$\le \bar{\rho}\le 2.57$ g/cm$^3$ is well known, while
the LPT2, near 175-2 g/cm$^3$ and LPT3 near 1.78-3 g/cm$3$ are proposed. The
inset shows metastable extensions of $F_{12}$  branches into the LPT2.5
region.  DFT-MD finds a nonmetallic LDL  in this region.}
\label{F12.fig}
\end{figure}

The total free energy $F=F_e+F_i+F_{emb}+F_{12}$, is  discussed  in the SM.
The $F_{12}$ term contains ion-ion bonding effects. Its
discontinuities indicate LPTs. The other terms
vary fairly smoothly with density. Figure~\ref{F12.fig} displays $F_{12}(\bar{\rho})$
for $l$-Si at 1200K. Similar (weaker) discontinuities are found even at
$T$ =1 eV (see SM). The spherically symmetric  model used 
allows only uniform-density solutions.
In the following, and in the SM, we discuss  the nature of the LPTs,
their PDFs,  the  compressibility $\kappa_T$ and the
electrical conductivity $\sigma$ across  them. Optical probes
can access  $\sigma$  and provide evidence for their onset.

\begin{figure}[t]                  
\includegraphics[width=0.95\columnwidth]{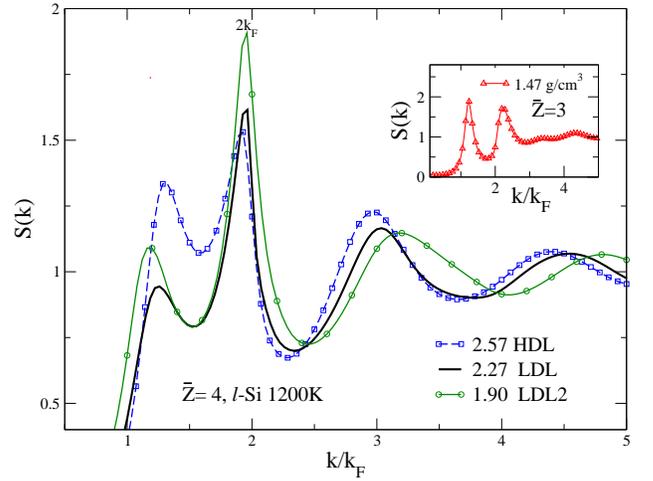}
 \caption
{(online color) The structure factor $S(k)$ of  LPT2.5 and LPT2
at 1200K. The main split-of peak
 registers at 2$k_F$ as required by the Fermi liquid picture
(this holds for LPT3 also, though not displayed).
 The x-axis is $k/k_F$ with $k_F$
appropriate to $\bar{Z}=4$. The inset for $S(k)$ at 1.47 g/cm$^3$,
$\bar{Z}$=3, is  beyond the IDT, and shows no 2$k_F$ splitting.}
\label{skDen.fig}
\end{figure}

{\it Discussion of structure data}-- The {\it average} charge state of the
 ion,  i.e., $\bar{Z}$,  is indirectly
 accessible from $N$-center DFT-MD simulations. 
 It can be measured via  the optical conductivity,
 X-ray Thomson scattering~\cite{GlenRed09},  or via  Langmuir probes. In
 a mixture of charge states, $\bar{Z}$ is the mean value  over
 the composition fractions $x_j$ of the integral
 charge states $Z_j$ (\cite{eos95}).
 This is the case for $\bar{\rho} < 1.5$ g/cm$^3$ when $Z_j$ may be
 4, 3, and 2 (see Fig.~\ref{DenZbar.fig}). 

Figure~\ref{DenZbar.fig} displays $\bar{Z}$
versus the free electron density $n_f=\bar{n}$. The IDT
and the pressure are further discussed in SM~\cite{supmat}).
As the NPA-HNC converges poorly near the IDT at 1200K, the 1 eV $P$-isotherm
is given. At 1200K we recover the Ganesh-Widom HDL 
pressure~\cite{,GaneshSiLPPT-09} extended up to LPT3. The PDFs at the IDT (discussed in SM) show
 short Si-Si {\it tr-cbs} with $r_b$ of $\sim$2.13 \AA \
 compared to $r_b$ in LDL-Si near 2.27-2.29/cm$^3$ at 1200K.  Near the IDT
 the Si-Si $r_b$ is 9\% shorter than in the solid, with stronger
bonding due to weaker screening for $\bar{Z}=3$.
 
Figure~\ref{skDen.fig} displays the variation of the $l$-Si $S(k)$ at 1200K,
 for $\bar{\rho}$ at 2.57 g/cm$^3$ (HDL), through the
  LPT2.5 to 2.27 g/cm$^3$ (LDL) and to 1.9 g/cm$^3$ at the LPT2.
 The major  peak in $S(k)$ is found to be  at $\sim2k_F$ as
expected from the FLP, with a subsidiary peak in the low-$k$ region. At
1.47 g/cm$^3$, beyond the IDT (inset, Fig.~\ref{skDen.fig}), $\bar{Z}=3$ and
no 2$k_F$ splitting exists.
 When transient
 bonding occurs, the valence $\bar{Z}$, a static average, does not change.
 Instead, the self-energy correction
 from 2$k_F$ scattering produces an
 increased  electron-effective mass $m^*$, with $m^*\simeq 1.1$. 
If $m^*=1$ the double peaks  of $S(k)$ become
a single peak at 2$k_F$ (see in SM~\cite{supmat}). This
 links the split peak, {\it tr-cbs} and the LPT with 2$k_F$ scattering
(see~\cite{utah12, DWP-carb90}).

\begin{figure}[t]     
\includegraphics[ width=0.98\columnwidth]{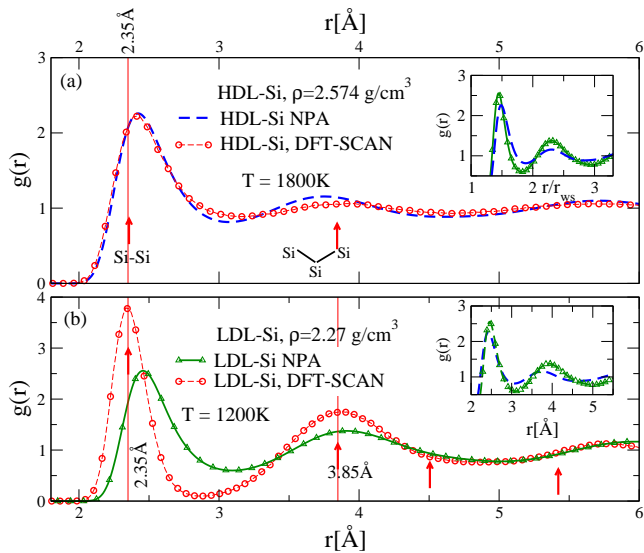}
 \caption
{(online color) $g(r)$ of (a) HDL (2.57g/cm$^3$) Si from NPA and from DFT-SCAN.
at 1800K, (b) LDL ($\sim$ 2.27g/cm$^3$) Si from NPA and DFT-SCAN.
The Si-Si bond length (2.35\AA ) is indicated.
The NPA-HNC fails to capture
 the increased  $g(r)$ at 2.35 \AA \,and remain locked in an HDL-like structure. 
Insets.  The NPA  $g(r)$, for HDL-Si and LDL-Si versus $r/r_{ws}$, and
 with $r$ in \AA .The HDL $r_{ws}\simeq 3$ a.u.}
\label{grHLsi.fig}
\end{figure}

The  PDFs  for the LDL and HDL reported in Ref.~\cite{Remsing18} are displayed in
~Fig.~\ref{grHLsi.fig} while their  $S(k)$ are displayed in the SM.
 Panel (a) of  Fig.~\ref{grHLsi.fig} shows the first peak of the NPA $g(r)$
 at 2.43 \AA . This is not determined by packing effects,
 but by {\it tr-cb}.  A  hard-sphere bridge term
 to the HNC equation via the  Lado-Foils-Ashcroft
 (LFA) criterion~\cite{LFA83} returns a
 negligible correction, as with $l$-C and $l$-Ge. 
 The HDL and LDL $g(r)$  show a 
 sharp  peak near 2.45 \AA \  some 4\% larger than the nominal Si-Si $r_b$  in
 the solid. The higher electron density in the liquid  weakens the Si-Si
interaction and leads to a longer average {\it tr-cb}.

The higher-$r$ peaks in the $g(r)$ of the HDL from DFT-SCAN studies  and from
 the NPA agree well, both for  peak position and height. 
As discussed in the SM,  Both $S(k)$ and $g(r)$ for
HDL-Si  from  DFT-SCAN   agree with NPA results, 
exhibiting the split-structure of
 the first subpeak of $S(k)$, while the higher-$k$  subpeak falls at $\sim2k_F$. 
The low-$k$ subpeak at $\sim 2$ \AA$^{-1}$ registers with the low-$k$
structure in the crystal $S(k)$, especially for the 
DFT-SCAN result for LDL-Si at 1200K.

In contrast, in panel (b),  the $g(r)$ for LDL-Si obtained
 from NPA and from DFT-SCAN  differ. A massive,  wide
1st peak at the Si nearest-neighbour (NN) position is observed in DFT-SCAN,
 with the second peak
 squarely at the  Si next-NN position. The NPA calculation for LDL-Si
 returns a $g(r)$ only slightly modified from HDL-Si at 1800K,
 as seen in the two insets, where the NPA $g(r)$  for the HDL and LDL
 at 1800K and 1200K are compared. The NPA-LDL is a metallic liquid, while
the LDL of DFT-SCAN is non-metallic (see conductivity).
%

\begin{figure}[t]
\begin{turn}{-90}
\includegraphics[width=0.75\columnwidth]{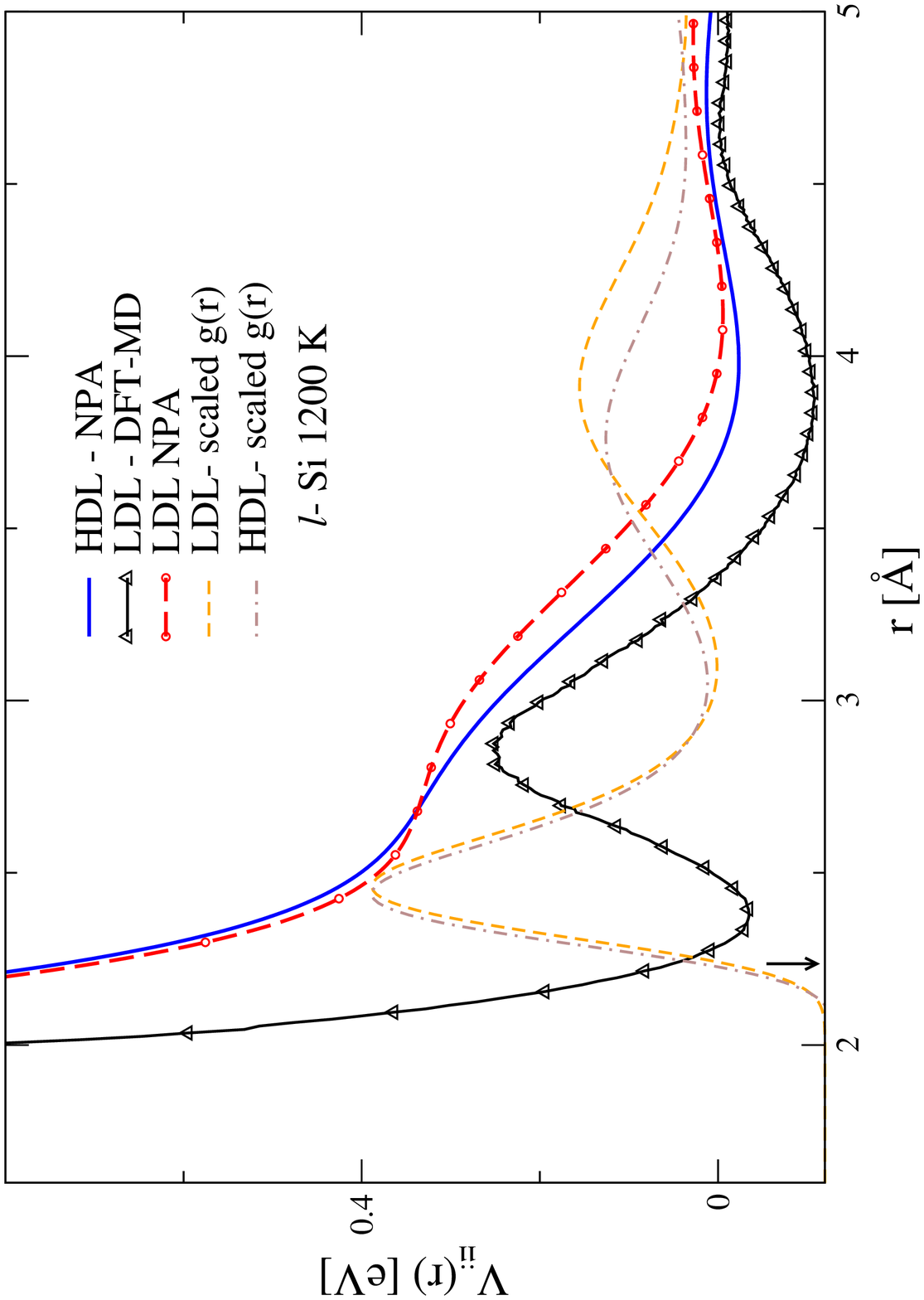}
\end{turn}
 \caption
{(online color)The ion-ion potentials  of the HDL and LDL are shown
against their PDFs (scaled to match  the potentials). The arrow at $r_b$
marks the Si-Si bond length. The first shell of ions sits on a repulsive ledge stabilized
by the electron cohesive energy, while the 2nd shell is in a Friedel-oscillation minimum (FOM).
Th HDL $\to$ LDL involves a cooperative movement of ions to the second shell as seen in
the enhanced LDL-$g(r)$ in the FOM region. This picture holds also for the LPT2 and LPT3 
(see SM). The DFT-SCAN potential accommodates the 1st shell at
its true minimum.}
\label{v-lpt.fig}
\end{figure}

To understand the temperature-robust LPTs found via the NPA,
and the  metastable LDL found so far via DFT-SCAN, we look at
 pair-potentials  in relation
to their PDFs (Fig.~\ref{v-lpt.fig}). The figure implies  a first-shell of ions on
a positive-energy ledge as also found  in liquid-aluminum~\cite{DWP-carb90}.
 These LDLs of  NPA-HNC  are  distinct  phases 
 separated from the HDLs by  free-energy jumps. With the coordination number
$N_c=6$ at the LPT2.5, the densities at LPT2 and LPT3 correspond to $N_c=5$ and $7$.
One may expect similar transitions at higher densities, until $N_c=12$ in a
high density solid form. The transitions are cooperative because the
Fermi length (1/$k_F$) is tied to the density via the peak of $S(k)$. 
 The HDL-LDL transition to $nm$-LDL found in DFT-MD is likely to be
 $N_c=6\to5$ and 4.
The metallic LDLs may be stable precursors of  metastable low-$T$ $nm$-LDLs,
as seen at LPT2.5, and likely to be seen at LPT2 as well. The inset to Fig.~\ref{F12.fig}
 shows metastable extensions of $F_{12}$ near the solid density.
Simulations that emphasize bonding and solid-like boundary conditions
succeed in detecting an  $nm$-LDL below the solid density here.

%
%

%
\begin{figure}[t]
\includegraphics[width=0.75\columnwidth]{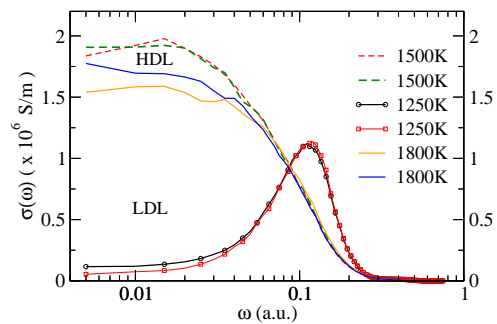}
 \caption
{(online color)  $\sigma(\omega)$ from DFT-MG-KG.
 for  and HDL  and LDL Si at LPT2.5. The LDL has a striking finite $\sigma(\omega)$,
 but the static $\sigma$ indicates a nearly non-metallic LDL.}
\label{sigOMG.fig}
\end{figure}

\begin{figure}[t]
\begin{center}
\includegraphics[width=0.95\columnwidth]{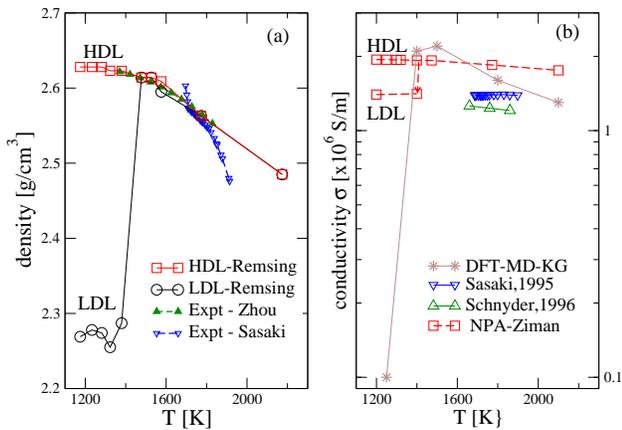}
 \caption
{(Online color)(a) The density $\bar{\rho}$ versus $T$ from DFT-SCAN ~\cite{Remsing18}
  near  LPT2.5 is compared with available
experimental densities
  (b) The static
conductivity $\sigma$  from 
NPA-Ziman, DFT-KG  and  from experiments~\cite{Schnyder96,Sasaki95}.
}
\label{condDen.fig}
\end{center}
\end{figure}

{\it Conductivity--}
The static electrical conductivity $\sigma=\sigma(\omega=0)$ is
measurable by `older' methods and via laser pump-probe
techniques.  In the degenerate regime ($T\ll E_F$)
electron scattering is
 mostly from $-k_F$ to $+k_F$, with a momentum transfer of
 $q\simeq 2k_F$. Even at
the highest $T$ studied here (1 eV), the electrons in $l$-Si are strongly
degenerate as $T/E_F\sim 0.078$ prior to the IDT. 
 As $T$ increases,  $\sigma$ decreases essentially linearly for  metals, though
not so in  $l$-Si data.This is reflected in our NPA-Ziman results
given in the SM. 
DFT-KG calculations yield
 a dynamic conductivity $\sigma(\omega)$ averaged over fixed
 configurations.  Fig.~\ref{sigOMG.fig} shows
 $\sigma(\omega)$ from typical DFT-KG calculations.
 The  difference in
the HDL  and LDL $\sigma(\omega)$ is striking. 
For $\omega<\omega(\sigma_{max})\simeq3$eV,
 the electrons seem to be in localized states of what may
 be a ``mobility edge'' of LDL-Si. 
 In HDL  {\it tr-cb}  the static limit
 $\omega\to0$ is metallic.
  
Fig.~\ref{condDen.fig} indicates that DFT-SCAN estimates of $\bar{\rho}(T)$ 
agree well with the experimental data of Zhou
 {\it et al }~\cite{ZhouMuk03},
but less so with  Sasaki {\it et al}~\cite{Sasaki95}.
Even when the density agrees,
 the $\sigma(0)$ from DFT-KG  (Fig.~\ref{sigOMG.fig}) or the NPA-Ziman
 (Fig.~\ref{condDen.fig}) are only in partial agreement with experiments.
Finite-$T$ $\sigma$ data at 1 eV are given in the SM.
The NPA treats  $l$-Si as a  single  fluid
 whereas many structures and  a distribution of
$m^*$ may be need to estimate $\sigma$.

{\it The IDT and critical-point models}--Discussions of the HDL-LDL phases
 in tetrahedral fluids have used a critical-point free
model~\cite{AngelWater14}, models with a liquid-liquid critical  point~\cite{ZhaoLqdSi16}, 
and models with two measures of order~\cite{EringDebene01}. The electron fluid
 and the FLP have  not figured much in these
discussions inspired from theories on non-metallic liquids like water.
The LPTs found in this study are most easily understood within the FLP.

{\it Conclusion}--
Liquid Si shows three liquid-liquid phase transitions and
possibly an  ionization-driven transition changing
$\bar{Z}=4$ to 3 in the density range 1.5-3 g/cm$^3$. The LPTs are linked to the
splitting of the main peak of the structure factor by concerted electron
scattering at the Fermi energy. The LPTs are robust and  are seen
(though  weakened) even at 1 eV. While the HDL near 2.5g/cm$^3$ found in this study
agrees with previous DFT-MD studies, the LDLs found via the NPA-HNC are stable metallic
 liquids  that may be
precursors to metastable non-metallic LDLs found in supercooled liquid Si near
2.27 g/cm$^3$. The HDL and LDL  conductivities
differ sufficiently and may provide an experimental signature of the LPTs
in optical and conductivity experiments~\cite{McMillan05}.
The compressibility (see SM)  displays
robust signatures of the LPTs, but they need to be confirmed
by  refined shock-Hugoniot type
experiments.
At densities 1.5 g/cm$^3$ the  free electron density is  reduced by localizing electrons
into the atomic core, leading to a possible IDT.

\end{document}